\documentclass{svproc}
\usepackage{amsmath,latexsym,amssymb}
\usepackage{url,graphicx}

\begin{document}
\mainmatter              % start of a contribution
\title{Parametric resonance of complex scalar field under spacetime oscillations}
\titlerunning{Parametric resonance of complex scalar field under spacetime oscillations}  % abbreviated title (for running head)
\author{Shreyansh S. Dave\inst{1} \and Sanatan Digal\inst{1,2}}
\authorrunning{Shreyansh S. Dave et al.} % abbreviated author list (for running head)
\tocauthor{Shreyansh S. Dave, Sanatan Digal}
\institute{The Institute of Mathematical Sciences, Chennai 600113, India,\\
\email{shreyanshsd@imsc.res.in},
\and
Homi Bhabha National Institute, Training School Complex, Anushakti Nagar, Mumbai 400085, India.}

\maketitle              % typeset the title of the contribution

\begin{abstract}
In this proceeding, we study time evolution of a complex scalar field, 
in symmetry broken phase, in presence of oscillating spacetime 
metric background. We show that spacetime oscillations lead to parametric 
resonance of the field. This generates excitations in the field for a wide 
range of frequency of spacetime oscillations which ultimately lead to the 
formation of topological vortices. The lowest frequency cut-off to induce 
this phenomena is set by system size due to finite size effects.  
\keywords{parametric resonance of classical field, topological vortices}
\end{abstract}
\section{Introduction}
There are various systems ranging from condesed matter physics to the early Universe 
where topological defects can exist and form under various conditions \cite{rev1}. 
They exist when the order parameter space or vacuum manifold of the system has a 
non-trivial topology \cite{mermin}. Formation of topological defects during symmetry 
breaking phase transitions are studied by Kibble-Zurek mechanism \cite{rev1,kbb}. 
Nucleation of superfluid vortex lattice in rotating vessel, flux-tube lattice in 
presence of magnetic field in type-II superconductor, etc. are other methods of 
formation of topological defects \cite{telley}.

The phenomena of parametric resonance of field, in symmetry broken phase, can also
produce topological defects \cite{reso,kink,sto}. This phenomena has been studied 
for periodically oscillating temperature of heat-bath, which generates excitations 
in the field leading to the formation of topological defects \cite{reso,kink}. In 
ref.\cite{sto}, we have shown that spacetime oscillations can also induce such 
phenomena for a wide range of frequencies. At frequencies lesser than the mass of 
field mainly transverse excitation of the field arises, while at higher frequencies 
longitudinal excitation also gets generated dominantly. In this case, the lowest 
frequency cut-off to generate field excitation is set by system size due to finite 
size effects. In this proceeding, we present some of the results of our work in 
ref.\cite{sto}.

\section{Equation of Motion}

For simplicity, we take the {\it inverse} spacetime metric as, 
$g^{\mu \nu}$$\equiv$$diag($$-$$1,1$$-$$h,1$$+$$h,1)$, where 
$h$$\equiv$$h(t,z)$=$\varepsilon \sin \big(\omega (t$$-$$z)\big)$;
($t,x,y,z$) are spacetime coordinates. The action of a complex scalar 
field on this spacetime manifold is given by \cite{sto},
\begin{equation}
 S=\int d^4x \sqrt{-g}\Big[-\frac{1}{2}g^{\mu \nu} \partial_{\mu}\Phi^* \partial_{\nu} \Phi - V(\Phi^* \Phi)\Big],
\end{equation}
where, $g$=$det(g_{\mu \nu})$=$-$$($1$-$$h^2)^{-1}$,
$\Phi$=$\phi_1$+$i\phi_2$, $\Phi^*$=$\phi_1$$-$$i\phi_2$; $\phi_1$ and $\phi_2$ are real scalar 
fields. We consider symmetry breaking effective potential as,
\begin{equation}
 V(\Phi^* \Phi)= \frac{\lambda}{4} \Big(\Phi^* \Phi - \Phi_0^2 \Big)^2,
\end{equation}
where, the mass of longitudinal-mode of field is $m_\Phi$=$\Phi_0 \sqrt{2\lambda}$. 
The equation of motion for ($\phi_1,\phi_2$) fields are \cite{sto},
\begin{equation}
 \Box \phi_i - \frac{dV}{d\phi_i}=0;~~
 \Box \phi_i = \frac{1}{\sqrt{-g}}\partial_{\mu}\Big(\sqrt{-g} g^{\mu \nu} \partial_{\nu}\phi_i \Big);~~i=1,2. 
\end{equation}
For the simplicity of solving it numerically, (i) we assume that there is no variation of the 
field $\Phi$ along $z$-direction, and (ii) we look at the solution of the field only in the 
$z$=0 plane. With these simplifications, the above equations in the expanded form become 
\cite{sto},
\begin{equation}
 \begin{split}
  -\frac{1}{2} \frac{\varepsilon^2 \omega \sin (2\omega t)}{f(t)f(-t)} 
  \frac{\partial \phi_i}{\partial t}  -\frac{\partial^2\phi_i}{\partial t^2} + 
 f(t) \frac{\partial^2\phi_i}{\partial x^2}
 +f(-t)\frac{\partial^2\phi_i}{\partial y^2} 
  - \lambda \phi_i \Big(\phi_1^2+\phi_2^2-\Phi_0^2 \Big)=0,
\end{split}
\label{eom1}
\end{equation}
where, $f(t)$=$1$$-$$\varepsilon \sin (\omega t)$. These equations clearly 
indicate that spacetime oscillations can affect field evolution iff the initial 
field configuration has some fluctuations, which can naturally be present 
due to thermal and/or quantum fluctuations. The momentum of field-modes of 
initial field configuration get coupled with spacetime oscillations and by 
following resonace conditions undergo parametric resonant growths \cite{sto}. 
One can show that when $\omega$$<$$m_{\Phi}$, then mainly transverse excitation 
of the field arise, while when $\omega$$>$$m_{\Phi}$ along with transverse
excitation, longitudinal excitation also gets generated dominantly \cite{sto}.

\section{Simulation details and results}

In our simulations, we have considered only transverse fluctuations in the 
initial field configuration. With this initial field configuration, we solve 
Eq.\ref{eom1} by using second order Leapfrog method and by considering periodic 
boundary conditions along spatial directions. We use lattice spacing of 
$\Delta x$=$\Delta y$=0.01 $\Lambda$ and evolve with time spacing of 
$\Delta t$=0.005 $\Lambda$.   

\begin{figure}
\begin{center}
%\vspace{2.5cm}
\includegraphics[width=0.45\linewidth]{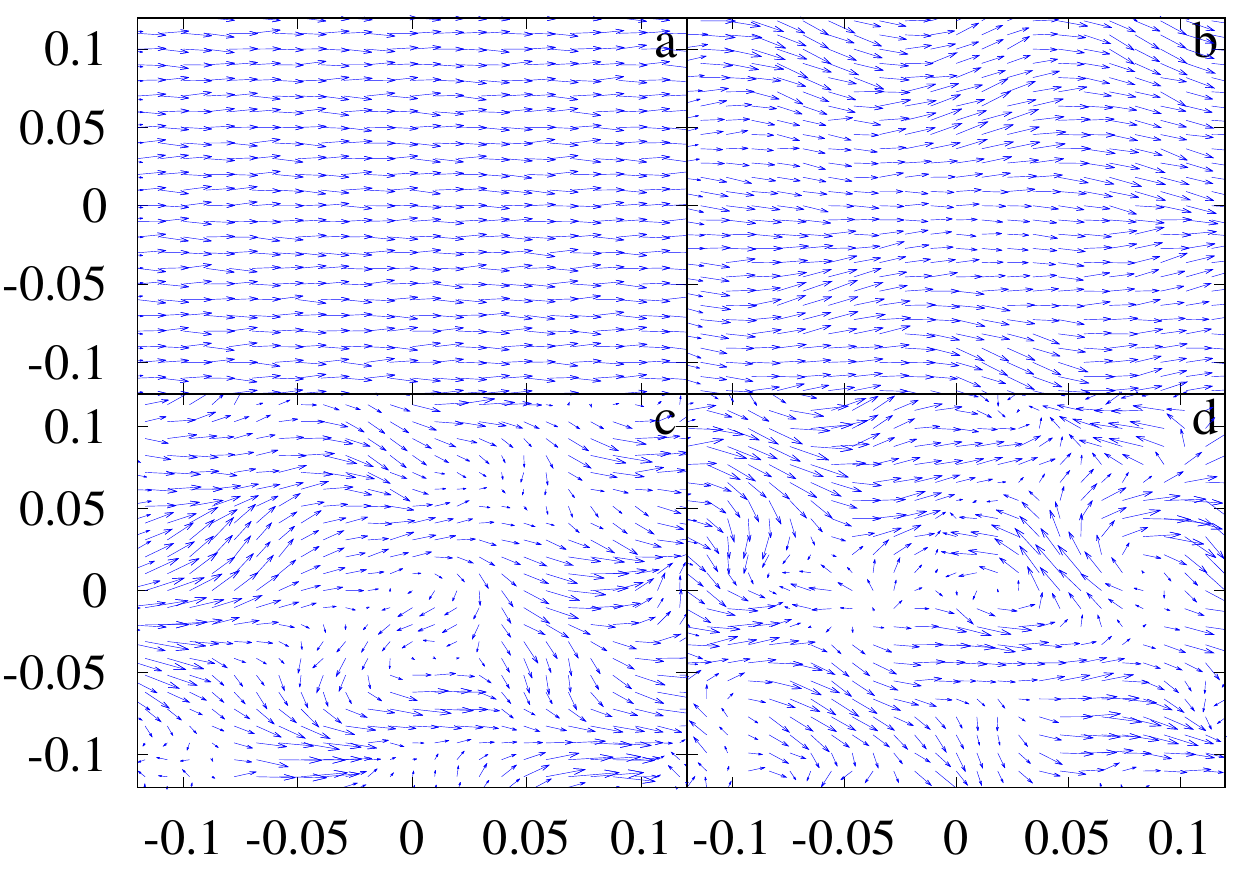}
\includegraphics[width=0.45\linewidth]{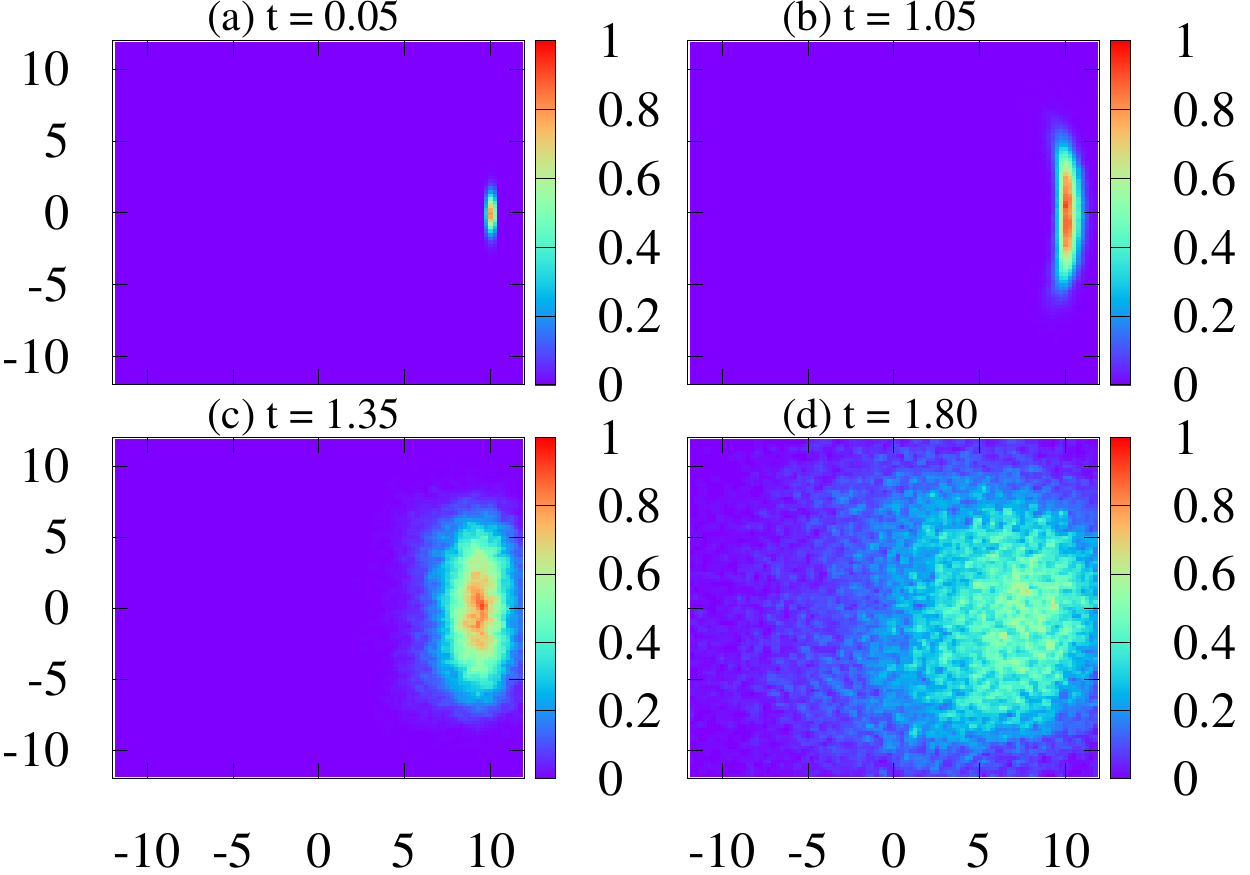}
\caption{Field configurations in physical space (left) and field distributions 
in field space (right) at different times of field evolution for $\omega$$>$$m_{\Phi}$. 
Figures are from ref.\cite{sto}.}
\label{fig1}
\end{center}
\end{figure}

\begin{figure}
\begin{center}
%\vspace{2.5cm}
\includegraphics[width=0.75\linewidth]{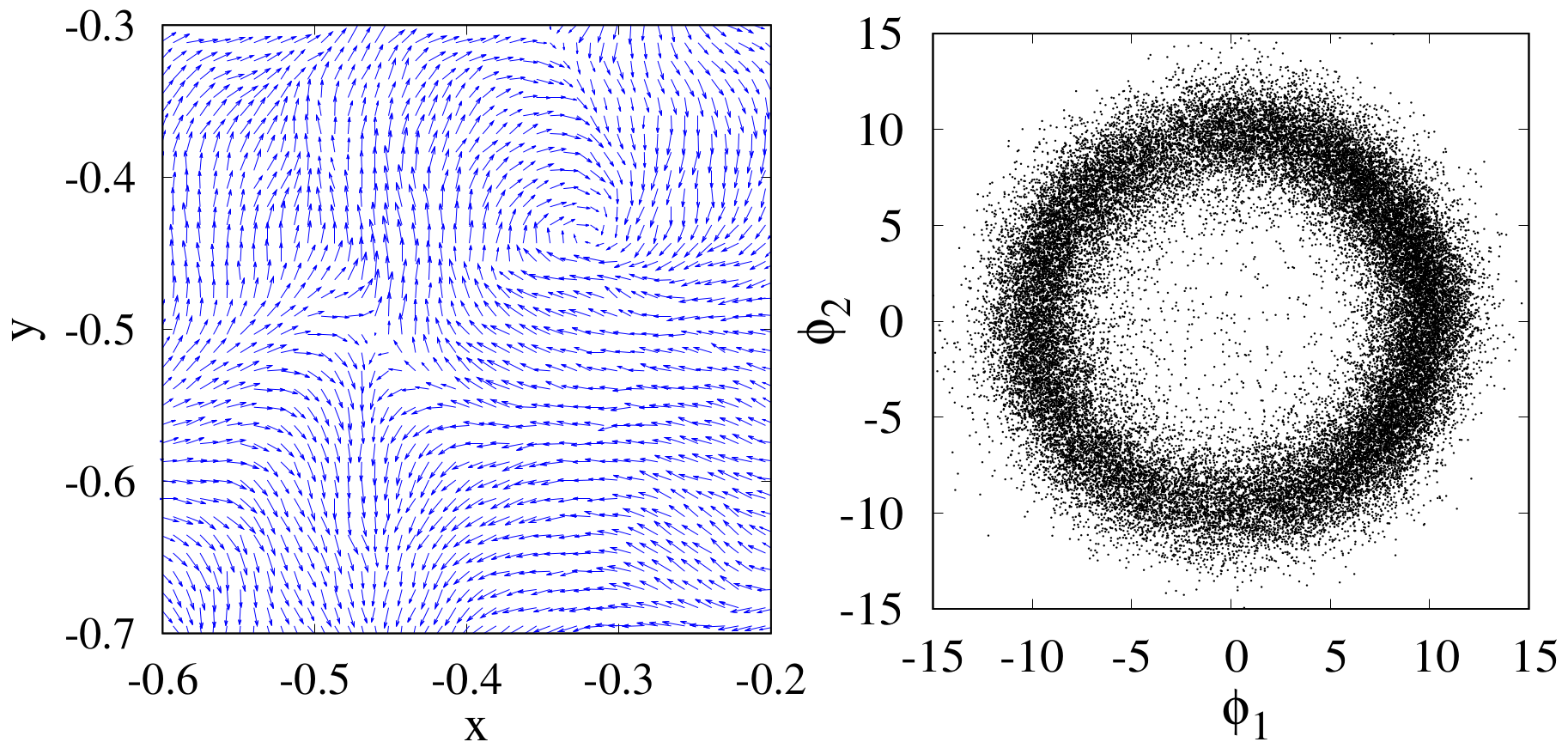}
\caption{Figures show formation of well separated vortices (left) and generation
of mainly transverse excitation in field space (right) for $\omega$$<$$m_{\Phi}$. 
Figures are from ref.\cite{sto}.}
\label{fig2}
\end{center}
\end{figure}

In Fig.\ref{fig1}, we have shown how the field gets excitations under spacetime 
oscillations in physical space (left) and in field space (right) at four different 
times of field evolution; Left: $t$=0, 1.35, 1.7, and 1.8 $\Lambda$, Right: $t$=0.05, 
1.05, 1.35, and 1.8 $\Lambda$. The parameters of simulations are $\varepsilon$=0.4, 
$\omega$=100 $\Lambda^{-1}$, $\Phi_0$=10 $\Lambda^{-1}$, and $\lambda$=40, which 
implies that for these parameters $\omega$$>$$m_{\Phi}$ allowing the dominant 
generation of longitudinal component of the field along with transverse component. 
Left plots clearly show that due to spacetime oscillations, the field generates a 
specific field-modes at the intermediate stage of the evolution. With further 
evolution, other field-modes also get generated which ultimately lead to the formation 
of topological vortices in the system. Right plots show that both component of the 
field, transverse as well as longitudinal, have been generated during the field 
evolution.
The generation of longitudinal excitation in this case makes the profile of the 
formed vortices highly distorted. In Fig.\ref{fig2}, we have taken $\omega$=20 
$\Lambda^{-1}$, for which $\omega$$<$$m_{\Phi}$ allowing the generation of mainly 
transverse excitation. This leads to the formation of well separated vortices. 
This has been depicted in physical space (left) and in field space (right) 
at time $t$=18.5 $\Lambda$. 

\begin{figure}
\begin{center}
%\vspace{2.5cm}
\includegraphics[width=0.45\linewidth]{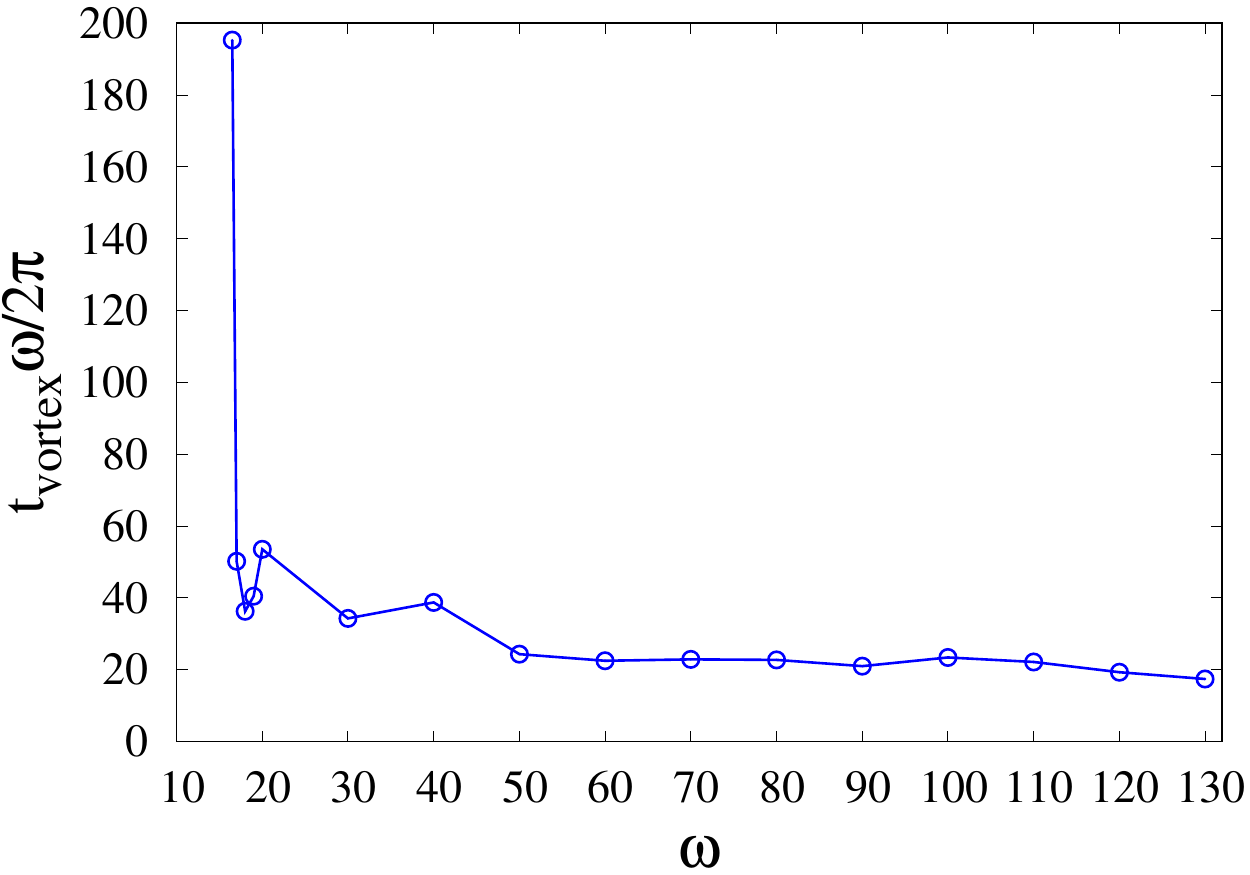}
\includegraphics[width=0.45\linewidth]{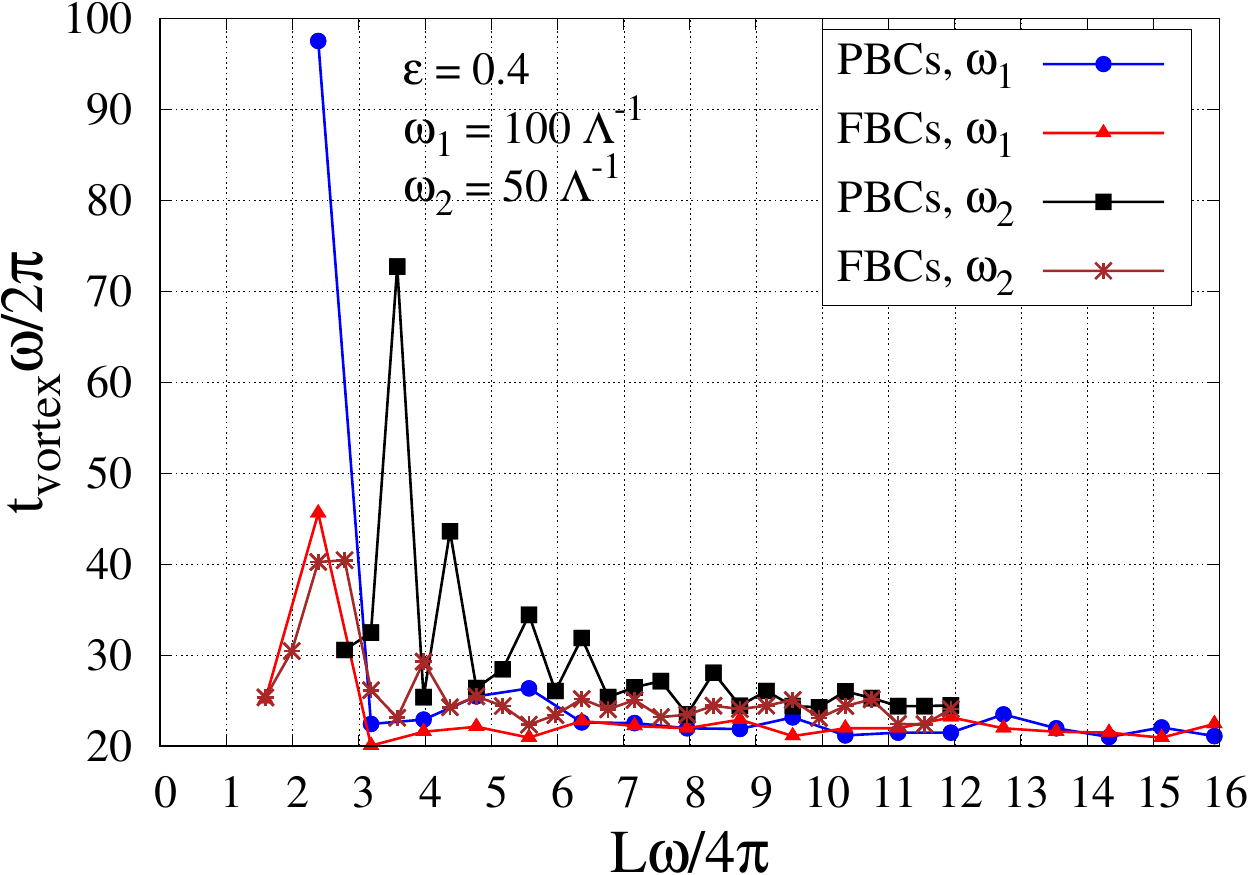}
\caption{Figures show response of $t_{vortex}$ to frequency $\omega$, choice of boundary 
conditions, and system size $L$. Figures are from ref.\cite{sto}.}
\label{fig3}
\end{center}
\end{figure}

In order to undestand the response of field to frequency $\omega$, we determine 
the time of formation of first vortex-antivortex pair in system and denote by 
$t_{vortex}$. A more useful quantity to analyse this response is 
$t_{vortex}~\omega/2\pi$, a dimensionless number, which counts the number of 
cycle (NoC) of spacetime oscillations up to $t_{vortex}$. In Fig.\ref{fig3} (left), 
we have plotted NoC versus $\omega$. This clearly shows that for sufficiently large 
$\omega$, NoC is almost independent from $\omega$. However, at low $\omega$, it 
starts deviating from such a constant value and diverges at very low $\omega$. This 
is an indication of finite size effects. In Fig.\ref{fig3} (right), we have studied 
the effects of using periodic and fix boundary conditions (PBCs and FBCs), system 
size $L$, and $\omega$ on NoC. For each curve, we have taken fixed value of $\omega$ 
($\omega_1$=50 $\Lambda^{-1}$ and $\omega_2$=100 $\Lambda^{-1}$), and $\varepsilon$=0.4, 
while we vary $L$. For sufficiently large values of $L\omega/4\pi$, NoC is independent 
from $L\omega/4\pi$ and takes roughly a constant value. On the other hand, for low 
$L\omega/4\pi$, it starts increasing and diverses when $L\omega/4\pi$$\sim$1. For PBCs 
these deviations are stronger than FBCs. Thus, the lowest frequency cut-off to induce 
this phenomenon is given by $\omega_{_L}$$\sim$$4\pi/L$. 



\begin{thebibliography}{6}
%

\bibitem{rev1}
Zurek, W.H.: 
Cosmological experiments in condensed matter systems.
Phys. Rep. 276, 177 (1996). \url{doi:10.1016/S0370-1573(96)00009-9}

\bibitem{mermin}
Mermin, N.D.: 
The topological theory of defects in ordered media.
Rev. Mod. Phys. 51, 591 (1979). \url{doi:10.1103/RevModPhys.51.591}

\bibitem{kbb}
Kibble, T.W.B.:
Topology of cosmic domains and strings.
J. Phys. A 9, 1387 (1976). \url{doi:10.1088/0305-4470/9/8/029}

\bibitem{telley}
Tilley, D.R., Tilley, J.: 
Superfluidity and Superconductivity.
Third Edition, Overseas Press (2005).

\bibitem{reso}
Digal, S., Ray, R., Sengupta, S., Srivastava, A.M.:
Resonant Production of Topological Defects.
Phys. Rev. Lett. 84, 826 (2000). \url{doi:10.1103/PhysRevLett.84.826}

\bibitem{kink}
Ray, R., Sengupta, S.:
Stochastic production of kink-antikink pairs in the presence of an oscillating background.
Phys. Rev. D 65, 063521 (2002). \url{doi:10.1103/PhysRevD.65.063521}

\bibitem{sto}
Dave, S.S., Digal, S.: 
Effects of oscillating spacetime metric background on a complex scalar field and formation 
of topological vortices.
arXiv: 1911.13216 [hep-th].

\end{thebibliography}
\end{document}